\documentclass[preprint,12pt]{elsarticle}

\usepackage{amssymb,amsmath,amsthm,booktabs}

\newcounter{bla}

\usepackage{graphicx,subcaption}
\usepackage[utf8]{inputenc} 
\usepackage[T1]{fontenc}
\usepackage[english]{babel}
\usepackage{aurical}

\addto\captionsenglish{}

\usepackage[centercolon=true]{mathtools} 
\usepackage{courier}
\usepackage{xspace,listings,microtype,suffix}
\usepackage[usenames,dvipsnames]{xcolor}
\usepackage[normalem]{ulem}
\usepackage{soul,calligra}

\usepackage{hyperref}
\usepackage[capitalize]{cleveref}
\hypersetup{colorlinks=true}

\def\tp#1{_{\text{#1}}}

\newcommand*{\DefaultAcronymStyle}[1]{\textsf{#1}}
\newcommand*{\Acronym}[2][\DefaultAcronymStyle]{\expandafter\def\csname #2\endcsname{#1{#2}\xspace}}
\WithSuffix\newcommand\Acronym*[3][\DefaultAcronymStyle]{\expandafter\def\csname #2\endcsname{#1{#3}\xspace}}

\newcommand*{\DefaultModuleStyle}[1]{\texttt{#1}}
\newcommand*{\Modul}[1]{\Acronym*[\DefaultModuleStyle]{#1}{#1}}
\WithSuffix\newcommand\Modul*[2]{\Acronym*[\DefaultModuleStyle]{#1}{#2}}

\Acronym{UNEW}
\Acronym{MATLAB}
\Acronym{Python}
\Acronym{YAML}
\Acronym{UWERR}
\Acronym{URL}

\Modul{unew}
\Modul{ioutils}
\Modul{analysis}
\Modul{session}
\Modul{plots}
\Modul{configuration}

\def\TT#1{\DefaultModuleStyle{#1}\xspace}

\journal{Computer Physics Communications}

\begin{document}

\begin{frontmatter}



\title{A Python program for the implementation of the $\Gamma$-method for Monte Carlo simulations}


\author[a,b]{Barbara De Palma}
\author[a,b]{Marco Erba\corref{author}}
\author[a,b]{Luca Mantovani}
\author[a,b]{Nicola Mosco}

\cortext[author] {Corresponding author.\\\textit{E-mail address:} marco.erba@unipv.it}
\address[a]{Dipartimento di Fisica, Universit\`a degli Studi di Pavia, Via A. Bassi 6, 27100, Pavia, Italy }
\address[b]{INFN, Sezione di Pavia, Via A. Bassi 6, 27100, Pavia, Italy\\
\begingroup
\normalfont
\bigskip\bigskip
\centering
\includegraphics[scale=0.1]{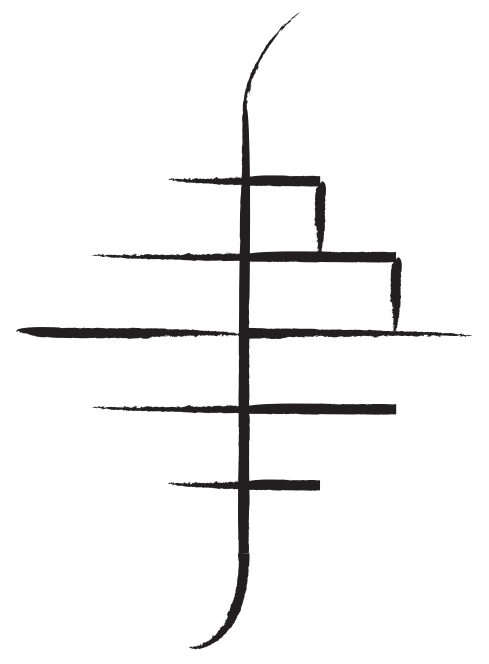}\\
\Fontskrivan
{Beto Collaboration Project}
\endgroup
}

\begin{abstract}
We present a modular analysis program written in \Python devoted to the estimation of autocorrelation times for Monte Carlo simulations by means of the $\Gamma$-method algorithm. We give a brief review of this method and describe the main features of the program. The latter is characterized by a user-friendly interface and an open source environment which, along with its modularity, make it a versatile tool. Finally we present a simple application as an operational test for the program.
\end{abstract}


\end{frontmatter}


\newpage


\noindent
{\bf PROGRAM SUMMARY}

\begin{small}
\noindent
{\em Program Title: UNEW}                                          \\
{\em Licensing provisions: MIT }                                   \\
{\em Programming language: Python }                                   \\
{\em Nature of problem: Computation of autocorrelation time for Monte Carlo generated data in an open source environment.
	}\\
{\em Solution method:
	Modular package implementing the $\Gamma$-method with advanced data handling.}
\end{small}

\section*{Introduction}

Monte Carlo (MC) simulations are nowadays an important and supportive tool for
various theoretical and experimental research areas.  A crucial aspect of
MC-simulations analysis is the accurate assessment of statistical and systematic
errors and the determination of the algorithm efficiency by means of the
computation of the autocorrelation time.  An effective way to address these
issues is given by the $\Gamma$-method \cite{ref1:LessErr}, a useful algorithm
allowing a better exploitation of the generated data with respect, for example,
to the usual binning techniques.  The $\Gamma$-method has been first implemented
in the \MATLAB function \UWERR\footnote{\MATLAB code for~\cite{ref1:LessErr}
available at \url{http://www.physik.hu-berlin.de/de/com/ALPHAsoft}.}.

In this work we present the \UNEW project, which consists in a \Python program
devoted to an analysis of MC data implementing the $\Gamma$-method.
The program has a modular structure that, combined with an open-source environment, allows for the possibility of extensions or embedding into preexisting software. Moreover, we provide various solutions
for input data handling with a user-friendly interface.

The paper is organized as follows. In Sec.~\ref{SecI} we review the $\Gamma$-method, defining all the
relevant estimators that the \UNEW program computes.  In Sec.~\ref{sec:unew} we
describe the general structure of \UNEW, referring to \ref{app:program} for the details of the
supporting \Python modules. In Sec.~\ref{SecIII} we present
a simple application of the program to test the correct implementation of the analysis
algorithm. In particular, we simulate the Ising model
with two different algorithms \cite{alg:metro,alg:cluster},
analyse the generated data and compare the results with the literature
\cite{alg:cl2}. Finally, we draw our comments and
conclusions in Sec.~\ref{SecIV}.

\section{Review of the analysis method}\label{SecI}

In this Section, we review the problem of the estimation of errors affecting
MC-generated data by means of the $\Gamma$-method developed by Wolff
\cite{ref1:LessErr}. This method relies on the evaluation the autocorrelation
present in the data, allowing for an assessment of the efficiency of the MC
algorithm employed.

\subsection{MC simulations and preliminary concepts}\label{SubSecIA}

Here we first sketch the main features of MC simulations, restricting our
attention to the field of \emph{Statistical Mechanics}. In this framework, one
is interested in simulating a physical system by sequentially generating an
ensemble $\{ q_i\}_{i=1}^{N}$ of configurations, sampled according to a given
distribution $P(q)$. The simulation is performed by means of an algorithm
characterized by an update rule which realizes a given transition $q \rightarrow
q'$ with probability $W(q \rightarrow q')$. We assume that the update process is a
Markov chain having $P$ as its unique equilibrium distribution (see
Ref.~\cite{Madras1988}), namely
\begin{equation*}
\sum_{q\in \mathcal{S}}P(q)W(q\rightarrow q') = P(q'),
\end{equation*}
where $\mathcal{S}$ is the space of all possible configurations. Furthermore, we also require that
\begin{equation}\label{eq:lim-W}
\lim_{n\to\infty} W^{(n)}(q\rightarrow q')=P(q'),
\end{equation}
where $W^{(n)}$ denotes the	 probability of a transition through $n$ update
steps.  In the following, we consider that the ensemble $\{ q_i\}_{i=1}^{N}$ is
generated after the Markov chain has reached equilibrium. From an operative
point of view, this is typically achieved by thermalizing the system for a
sufficiently long time.

For each configuration $q_i$, one evaluates a set $\{ A_\alpha
\}_{\alpha=1}^{N_\alpha}$ of physical observables, called \emph{primary
observables}, whose MC-evaluations are denoted as $\mathcal{O}_\alpha(q_i) =
a^i_\alpha$, for all $i=1,\ldots,N$. The output of the simulation can then be
arranged into a $N\times N_\alpha$ matrix. In general, one is interested in
evaluating also a \emph{derived quantity}, namely a function $f$ of the primary
observables which we denote as
\[
F = f(A_1,A_2,\ldots, A_{N_{\alpha}})=f(A_{\alpha}).
\]
In particular, this is necessary for those observables which can not be defined configuration by configuration.

Without loss of generality, in the following we will illustrate the basic concepts of the $\Gamma$-method referring only to the analysis of primary observables. A review of the method for the case of a generic function can be found in \ref{app:1}.
We remark that the analysis method
examined in this manuscript holds regardless of the algorithm used.

\subsection{Review of the $\Gamma$-method}\label{SubsecIB}

In general, data generated with MC algorithms exhibit \emph{autocorrelation}: given two different time-steps of the simulation $i$ and $j$,
the corresponding estimates $a_\alpha^{i}$ and $a_\alpha^{j}$ can not be
considered statistically uncorrelated.
The $\Gamma$-method provides an accurate assessment of the autocorrelation and thus of the error affecting correlated data.

The statistical correlation is captured by the
\emph{correlation matrix}, defined, under the equilibrium assumption, as
\begin{align*}
	\Gamma_{\alpha\beta}(n)
	& =
		\left\langle\left(a_{\alpha}^{i}-A_{\alpha}\right)
		\left(a_{\beta}^{i+n}-A_{\beta}\right)\right\rangle \\
	& = \sum_{q,q'\in \mathcal{S}} P(q) W^n(q\to q')
		(\mathcal{O}_\alpha(q) - A_\alpha)
		(\mathcal{O}_\beta(q') - A_\beta).
\end{align*}
The diagonal elements $\Gamma_\alpha\equiv\Gamma_{\alpha\alpha}$ are called
\emph{autocorrelation functions}.  We remark that, here and in the following,
averages indicated by the bracket notation $\langle \cdot\rangle$ are taken on
ensembles of identical numerical experiments with independent random numbers and
initial states.
Typically, the autocorrelation functions exhibit an exponential decay
for large times
\begin{equation}\label{eq:gammainf}
\Gamma_{\alpha}(n) \sim \exp \left( -\frac{n}{\tau} \right), \quad n\to\infty.
\end{equation}
In a broad variety of cases, the decay constant $\tau$ has the same order of
magnitude of the equilibration time: accordingly, as a rule of thumb, an
estimate of $\tau$ can be used to verify, \emph{a posteriori}, that the
thermalization time of the system was much bigger than $\tau$
\cite{ref1:LessErr,Madras1988}.

In order to compute the statistical error and to assess
the efficiency of the MC algorithm used to generate the data we define, for each observable $A_\alpha$, the so called \emph{integrated autocorrelation time}
\begin{align}
	\tau_{\text{int},\alpha} & =
		\frac{1}{2}\sum_{t=-\infty}^{\infty} \rho_{\alpha}(t) =
		\frac{1}{2} \frac{C_\alpha}{\Gamma_{\alpha}(0)}, \\
\shortintertext{where $\rho_\alpha(t)$ is the normalized autocorrelation function}
	\rho_\alpha(t) & =
		\frac{\Gamma_{\alpha}(t)}{\Gamma_{\alpha}(0)} \label{eq:rho} \\
\shortintertext{and $C_\alpha$ is the autocorrelation sum given by}
	C_\alpha & = \sum_{t=-\infty}^\infty\Gamma_\alpha(t). \label{eq:auto-sum}
\end{align}
The quantity $\tau_{\mathrm{int},\alpha}$ gives an estimate of the error of $A_\alpha$ due to the autocorrelation, once the Markov chain has
been equilibrated. If we consider the sample mean
\begin{align} \label{eq:mean}
	\bar a_{\alpha} = \frac{1}{N}\sum_{i=1}^N a_\alpha^i,
\end{align}
as an estimator of the exact value $A_\alpha$, it can be shown that the resulting error $\sigma_\alpha$ of $\bar a_{\alpha}$ is given by
\begin{align}\label{eq:sigma-a}
	\sigma_\alpha^2 \approx
		\frac{2\tau_{\text{int},\alpha}}{N}\Gamma_{\alpha}(0) \quad
		\text{for} \quad N \gg \tau.
\end{align}
Therefore the variance given by $\Gamma_{\alpha}(0)$ is modified by
a factor of $\frac{2\tau_{\text{int},\alpha}}{N}$ in presence of
autocorrelations.

A significant issue, that may also affect the interpretation of the analysis
results, concerns the practical estimate of the integrated autocorrelation time.
We first need to introduce the estimator of the autocorrelation function
associated to the observable $A_{\alpha}$:
\begin{equation} \label{eq:gamma-est}
\bar{\Gamma}_{\alpha}(t)=\frac{1}{N-|t|}\sum_{i=1}^{N-|t|}\left( a_{\alpha}^{i}-\bar{a}_{\alpha}\right)\left( a_{\alpha}^{i+t}-\bar{a}_{\alpha}\right).
\end{equation}
As a natural estimator for $\tau_{\text{int},\alpha}$ we could take
\begin{align}\label{eq:wrong-tau}
	&\bar\tau_{\text{int},\alpha}(N-1) =
	 	\frac{1}{2} \frac{\bar C_\alpha(N-1)}{\bar\Gamma_\alpha(0)}, \\
\shortintertext{where}
	&\bar C_\alpha(N) = \bar\Gamma_\alpha(0) + 2\sum_{t=1}^{N} \bar\Gamma_\alpha(t)
\end{align}
is the estimator for the autocorrelation sum~\eqref{eq:auto-sum}.  However, it
turns out that the variance of the estimator~\eqref{eq:wrong-tau} does not
vanish as $N$ goes to infinity~\cite{Madras1988,Priestley1981}, due to the
presence of noise in the tail of $\rho(t)$. For this reason, we need to
introduce a summation window $W<N$ into \cref{eq:wrong-tau}.  As a side effect,
such a truncation leads to a bias in the autocorrelation sum,
\begin{align}
	\left|\frac{\left\langle\bar C_\alpha(W)\right\rangle-C_\alpha}{C_\alpha}\right| \sim e^{-\frac{W}{\tau}},
\end{align}
which eventually translates into a systematic error associated to the error $\sigma_\alpha$ of
the observable $A_\alpha$.  Therefore, the choice of the summation window $W$
should be made with care: it has to be large enough compared to the decay time
$\tau$ so as to reduce the systematic error, but at the same time not too large
in order to avoid the inclusion of excessive noise.  We take as optimal the
summation window $W$ that minimizes the total relative error (sum of the
statistical and systematic errors) on the considered observable
\cite{ref1:LessErr}:
\begin{equation} \label{eq:rel-err}
	\frac{\delta_{\text{tot}}(\bar\sigma_\alpha)}{\bar\sigma_\alpha}\approx\frac{1}{2}\min_W\left(e^{-\frac{W}{\tau}}+2\sqrt{W/N}\right),
\end{equation}
where $\bar\sigma_\alpha^2=\bar C_\alpha(W)/N$. In practice, such a value of $W$ can be determined by using the automatic procedure proposed in Ref.~\cite{ref1:LessErr}. Under the assumption of an exponential decay of the autocorrelation function, we can write
\begin{equation} \label{eq:tauw}
2\bar\tau_{\text{int},\alpha}(W)=\sum_{t=-\infty}^{\infty}\exp\left(-\frac{S|t|}{\bar{\tau}(W)}\right)
\end{equation}
where $\bar\tau_{\text{int},\alpha}$ is defined in \cref{eq:wrong-tau}, $S$ is a positive factor and $\bar\tau(W)$ is an estimator for the decay rate $\tau$. The $S$ factor can be adjusted to account for possible discrepancies between $\tau$ and $\bar\tau(W)$. By inverting \cref{eq:tauw} one finds, at the first order, $\bar\tau(W)\sim S\bar\tau_{\text{int},\alpha}$: we use this value of $\bar\tau(W)$ to evaluate the minimum of \cref{eq:rel-err}, which yields the optimal value $W\tp{opt}$ for $W$ (see \ref{app:1} for more details).
As a consistency check of the resulting summation window, one can verify, by adjusting the value of $S$, that the plot of the integrated autocorrelation time as a function of $W$ exhibits a plateau around the optimal value.

We finally introduce a slight generalization of the framework presented above. The set of $N$ data can be divided into $R$ statistically
independent replica, which in turn may be produced by parallel simulations or by splitting the data produced by a single run. Each replicum contains $N_r$ estimates: we denote with $a_\alpha^{i,r}$ the $i$-th MC estimate of the $r$-th replicum.
The autocorrelation function satisfies
\begin{equation*}
\left\langle
\left(a_{\alpha}^{i,r}-A_{\alpha}\right)
\left(a_{\beta}^{i+n,s}-A_{\beta}\right)\right\rangle =
\delta_{rs}\Gamma_{\alpha\beta}(n).
\end{equation*}
Notice that $N_r$ must be chosen carefully, in order to effectively end up with
statistically independent replica; in particular, if $N_r \gg \tau$ does not
hold, the error estimation fails. The definition of the estimators for the
general case with $R>1$ is given in \ref{app:1}.

\section{Program and library}\label{sec:unew}

The main purpose of the \UNEW project is to provide a user-friendly interface to
the implementation of the $\Gamma$-method in an open-source environment. To this end, we
consider \Python to be the optimal language, since it features
a rich set of modules---from statistical and numerical scopes to simple yet
powerful graphics capabilities---and it is also widely used in academia.

The \UNEW project consists of a \Python package named \unew that can be run as a
command line tool and it contains a number of \Python modules serving as a
supporting library. The \unew package can be installed and used on every
platform provided with an installation of \Python 2.7 or higher, along with the
required packages. In this section, we will briefly illustrate the
structure and purpose of the modules of the package.

Input data are handled by the functions defined in the module \ioutils.  The
$\Gamma$-method is implemented in the module \analysis, which computes both the estimators of the mean values and the associated errors
in terms of the integrated autocorrelation time, as explained in~\cref{SecI}.
The classes defined in the module \plots manage all the necessary resources used
to plot the results of the analysis.  The package provides also the
\configuration module to acquire from the input parameters the necessary information
to execute the analysis.  The program is designed to handle an arbitrary number
of files (considered as different replica of the same experiment), each one with data arranged in a $N_k\times N_\alpha$ matrix, $N_k$ being the
number of rows of the $k$-th file.
Each file can be split into further segments that will be, in turn, treated as independent replica.

At the end of the analysis process, the \UNEW program returns the following output: the
mean value of the selected observable(s), its error, the error of the error, the
variance, the naive error computed disregarding autocorrelation, the integrated
autocorrelation time along with its error, and the optimal summation window.
Additionally, plots are produced for convenience of the user, showing the
integrated autocorrelation time as a function of the summation window, the
normalized autocorrelation, the histogram of replica and the distribution of the
data.

We stress again that the program is devised so that the analysis process applies
to both primary and derived observables. In \ref{app:program}, the reader can
find more details on the installation instructions and on the structure of
the package.


\section{Application}\label{SecIII}

In this section we test our implementation of the $\Gamma$-method, illustrating
an application to the well known Ising model. We consider a collection of spin variables arranged into a square lattice of size $L$ in absence of external magnetic fields, and simulate the model resorting to two different algorithms: \emph{Metropolis}~\cite{alg:metro} and the \emph{single cluster Wolff algorithm}~\cite{alg:cluster}.

The scaling of the autocorrelation time $\tau$ as a function of the \emph{correlation length} $\xi$ near the phase transition (i.e. for $\xi\to\infty$) reads
\begin{equation} \label{eq:z}
	\tau \sim \xi^z.
\end{equation}
The \emph{dynamic critical exponent} $z$ depends on the update rule
employed and, as one can see from \cref{eq:z}, it gives an assessment of the
efficiency of the algorithm. The practical estimation of $z$ can be done using
\begin{equation*}
	\tau_{\text{int},\alpha} \sim L^{z_\alpha}, \quad L\to \infty,
\end{equation*}
by fitting $\tau_{\text{int},\alpha}$ as a function of $L$ and extrapolating the
value of $z_\alpha$. As a consequence, one introduces a dependence of $z$ also
on the particular observable under consideration.

In our analysis, we focus on the following observables: i) the energy density $E$ and ii) the scaling quantity
\begin{equation}
\label{eq:fchi}
f_\chi = \dfrac{L^{7/4}}{\chi},
\end{equation}
where $\chi$ is the susceptibility.
We will discuss our results and compare them with those given in  Ref.~\cite{alg:cl2} and the reader can find in \ref{sec:example} a practical example of the program execution.

\subsection{Results}
For the test of the \UNEW implementation of the $\Gamma$-method, we produced the following statistics: $10^5$
thermalization steps and $10^7$ sweeps. Our results shall be compared with those of Ref.~\cite{alg:cl2}. The data have been generated with the
two aforementioned algorithms at the critical temperature $T_c = 1/\beta_c$,
where $\beta_c = \frac{\ln(1+\sqrt{2})}{2}$. As an operational test to check the
correct implementation of the $\Gamma$-method, we performed the analysis using
also the \MATLAB function \UWERR\footnote{\MATLAB code for~\cite{ref1:LessErr}
available at \url{http://www.physik.hu-berlin.de/de/com/ALPHAsoft}.}, obtaining
identical results.

\begin{table}[t]
	\centering
	\begin{tabular}{ccccc}
		$L$	& $f_\chi$     & $\tau_{f\chi}$   & $E$  & $\tau_{E} $    \\
		\midrule 
24  & 0.91723(28)    & 2.72(1)        & 0.720133(50) & 3.34(1)  \\
32  & 0.91649(30)    & 3.10(1)        & 0.716889(43) & 3.96(2)  \\
48  & 0.91613(32)    & 3.69(2)        & 0.713596(34) & 4.93(2)  \\
64  & 0.91574(34)    & 4.15(2)        & 0.711995(28) & 5.77(3)  \\
80  & 0.91605(36)    & 4.55(2)        & 0.710990(24) & 6.46(3)  \\
128 & 0.91633(39)    & 5.46(3)        & 0.709519(18) & 8.24(5)  \\
256 & 0.91555(45)    & 7.06(4)        & 0.708331(11) & 11.49(8)
	\end{tabular}
	\caption{Results obtained with the Wolff algorithm, performing $10^5$ thermalization steps and $10^7$ sweeps.}
	\protect\label{tab:SW}
\end{table}

\begin{table}[h!]
	\centering
	\begin{tabular}{ccccc}
		$L$	    & $f_\chi$     &  $\tau_{f\chi}$  & $E$  & $\tau_{E} $    \\
		\midrule 
24  & 0.91742(76)    & 20.4(2)        & 0.719938(93) & 11.25(8) \\
32  & 0.91864(103)   & 37.5(4)        & 0.716591(92) & 18.4(2)  \\
48  & 0.91463(154)   & 86(1)          & 0.713680(92) & 36.9(4)  \\
64  & 0.91816(218)   & 165(4)         & 0.711867(96) & 66(1)    \\
80  & 0.92071(269)   & 248(7)         & 0.710837(94) & 96(2)    \\
128 & 0.91769(448)   & 696(31)        & 0.709485(96) & 230(6)   \\
256 & 0.93102(974)   & 2967(249)      & 0.708189(99) & 858(42)
	\end{tabular}
	\caption{Results obtained with Metropolis algorithm, performing $10^5$ thermalization steps and $10^7$ sweeps.}
	\protect\label{tab:metro}
\end{table}

\begin{figure}[t]
	\centering
	\includegraphics[width=0.495\textwidth]{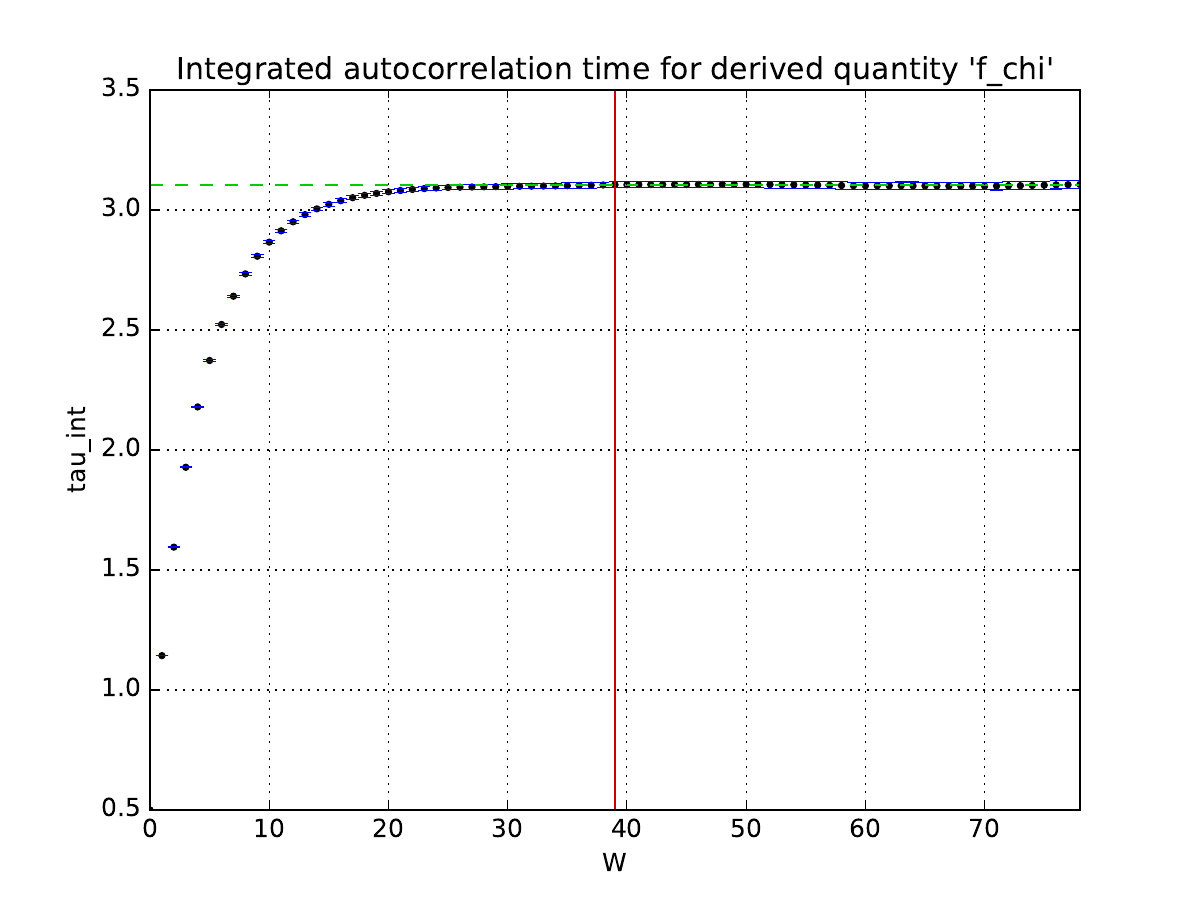}
	\includegraphics[width=0.495\textwidth]{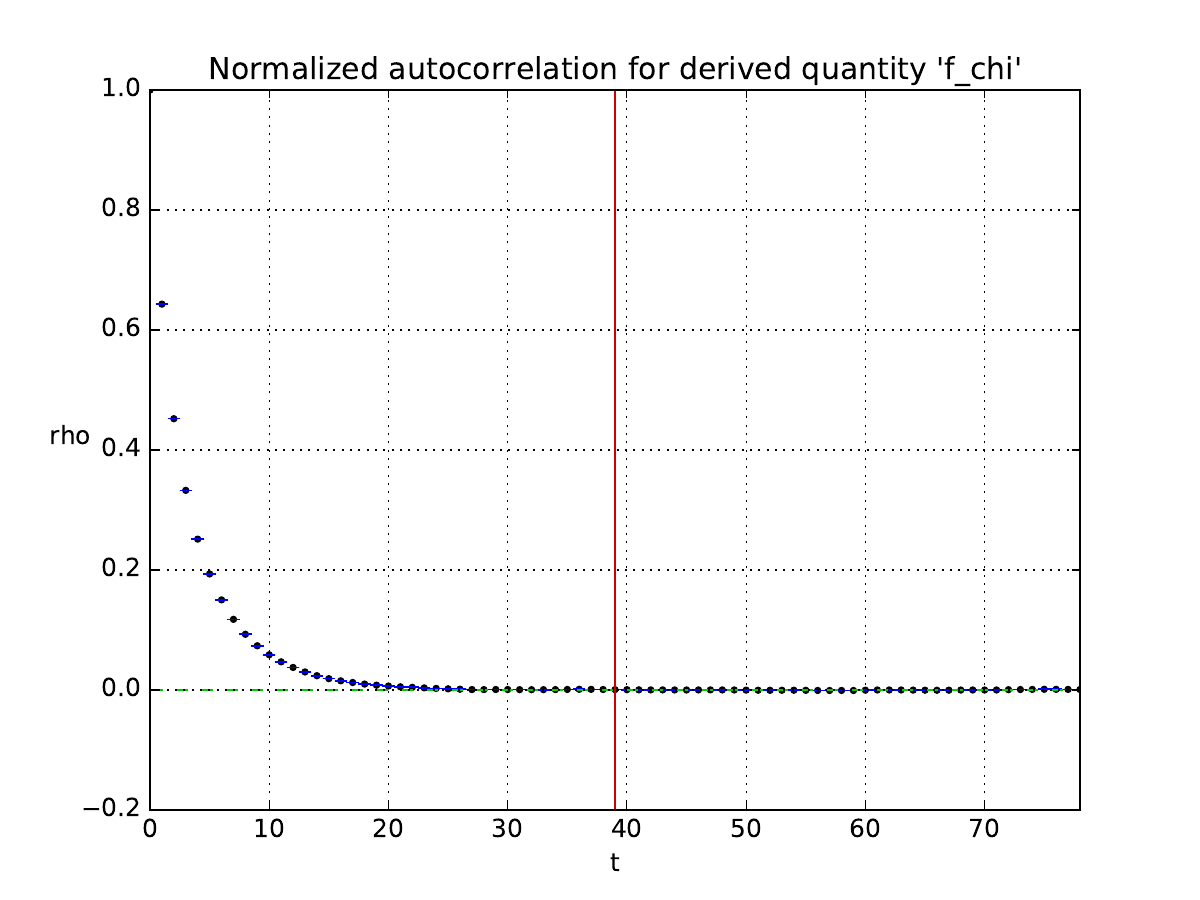}
	\\[0.4cm]
	\includegraphics[width=0.495\textwidth]{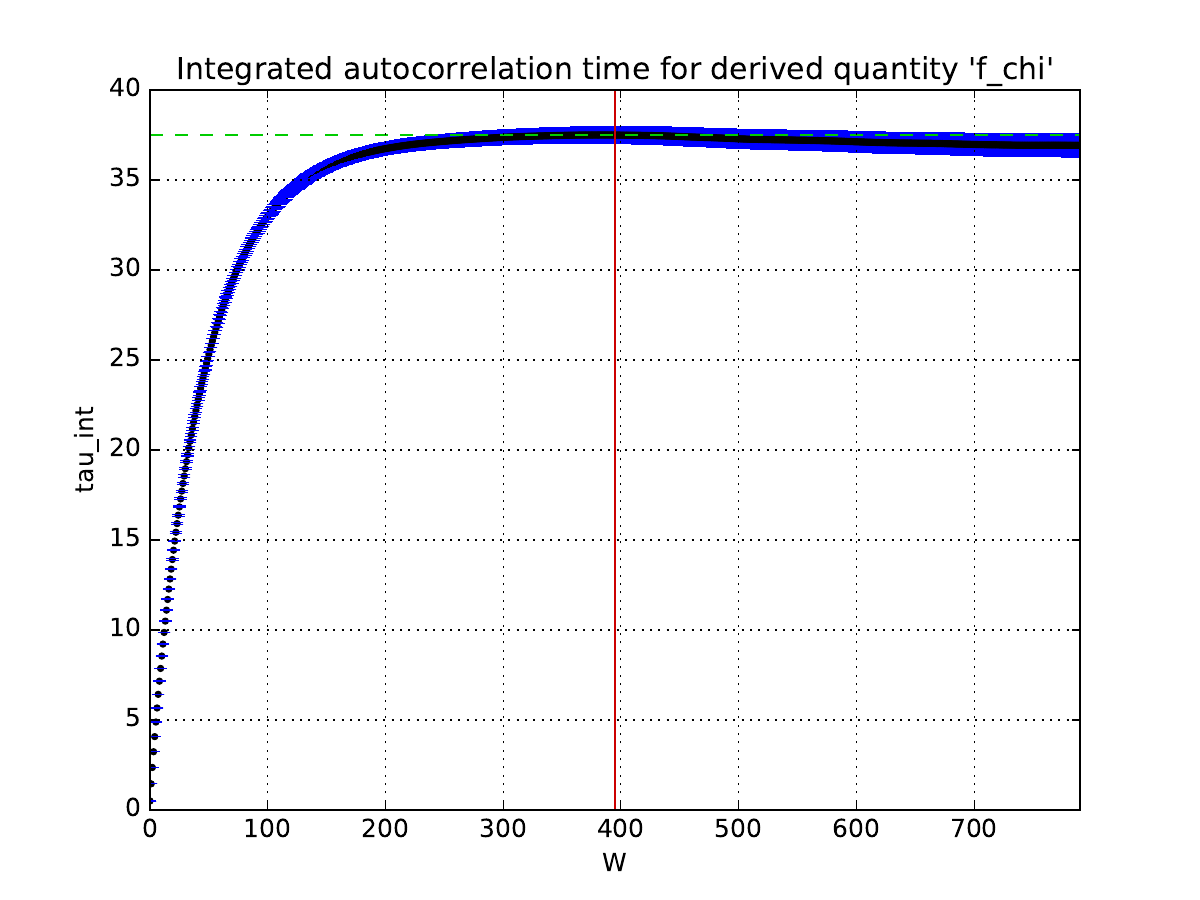}
	\includegraphics[width=0.495\textwidth]{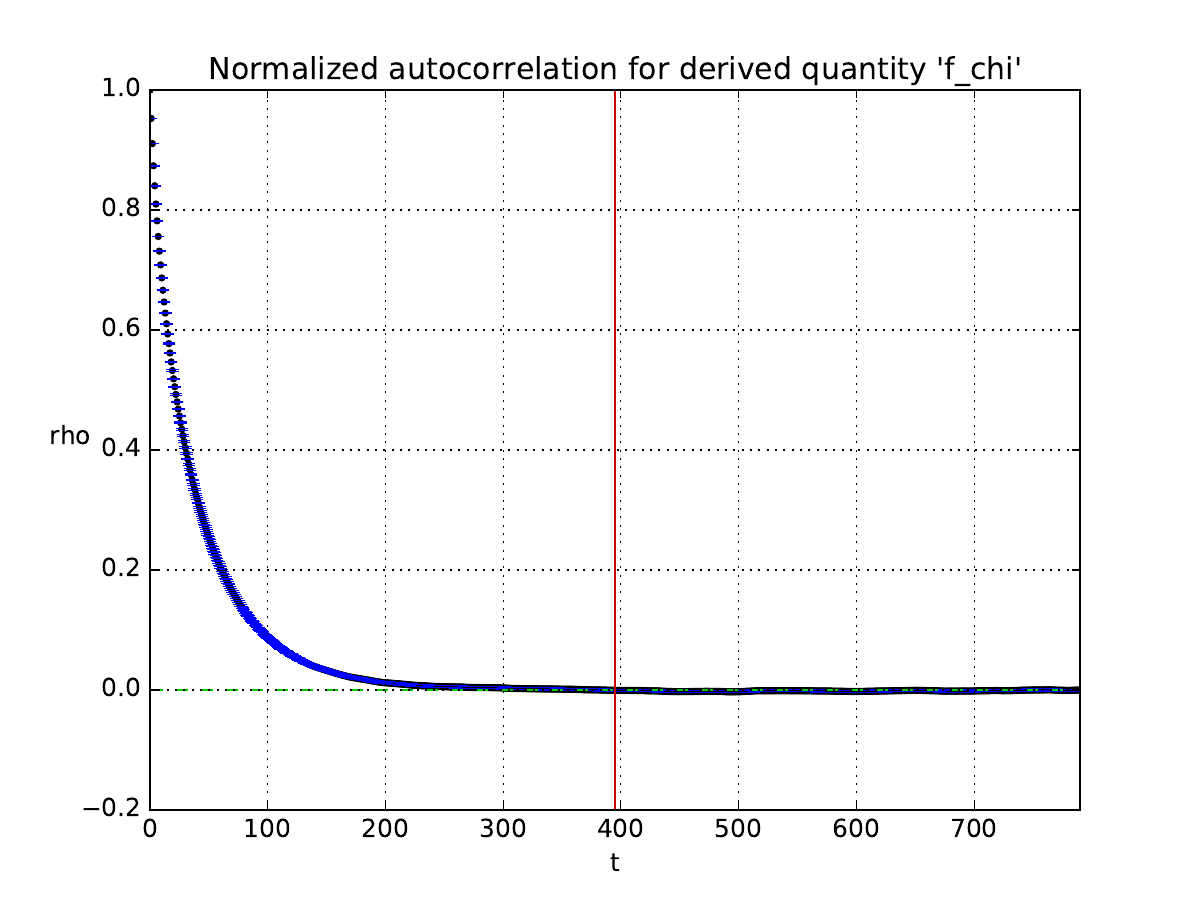}
  \caption{Integrated autocorrelation time $\tau_{f_\chi}$ as a function of the
  summation window W (left) and corresponding normalized autocorrelation $\rho$
  as a function of $t$ (right) for the single cluster and Metropolis algorithms
  (top to bottom), with $L=32$. The red vertical line corresponds to the
  optimal value of the summation window.}
	\label{fig:tau_rho}
\end{figure}

In \cref{tab:metro,tab:SW} we report the results of our analysis
respectively for the single cluster and Metropolis algorithm.  For each algorithm we consider different lattice sizes $L$ and evaluate $E$ and $f_\chi$ along with the respective autocorrelation times. Moreover, in \cref{fig:tau_rho} we show  the \UNEW plots of the autocorrelation time
and of the normalized autocorrelation for the derived quantity $f_\chi$ in the
case $L=32$.

Firstly, we notice that the simulations with the two different algorithms are in
agreement at the $2\sigma$--level for the mean values of $f_\chi$ and $E$. Furthermore, as expected, the behaviour of the autocorrelation times
(\cref{fig:tau_rho}) exhibits a plateau starting approximately from the optimal
summation window $W\tp{opt}$; at the same time, summing the autocorrelation
function up to $W\tp{opt}$ allows to exclude (at least) a significant portion of
its noisy tail. This has been achieved via a proper choice of the $S$ factor, as
discussed in \cref{SubsecIB}.

In a typical scenario a suitable choice for $S$ is around a few units
\cite{ref1:LessErr}: this is indeed what we find for both algorithms.

\begin{figure}[t]
	\centering
	\includegraphics[width=0.4955\textwidth]{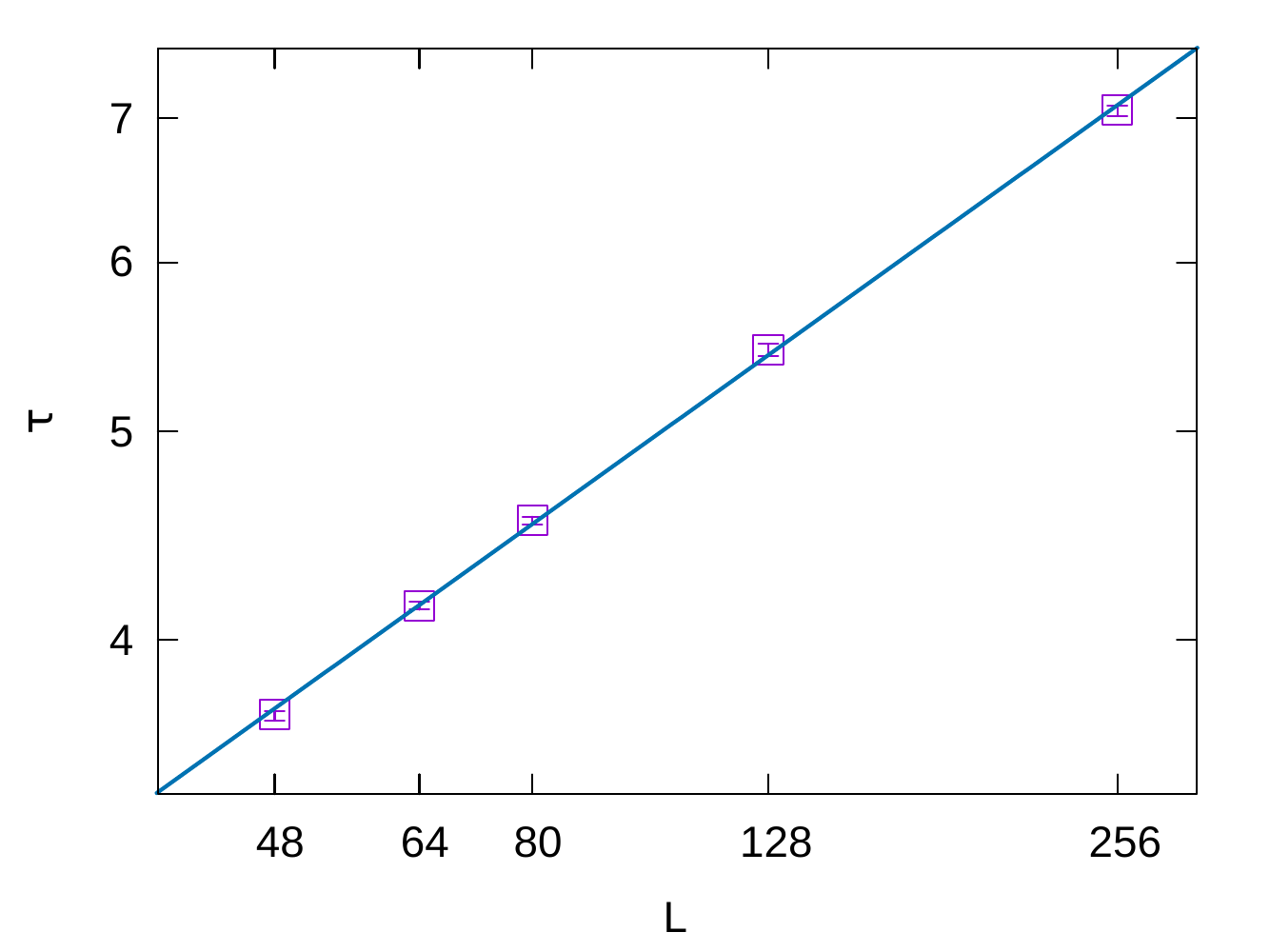}
	\hfill
	\includegraphics[width=0.4955\textwidth]{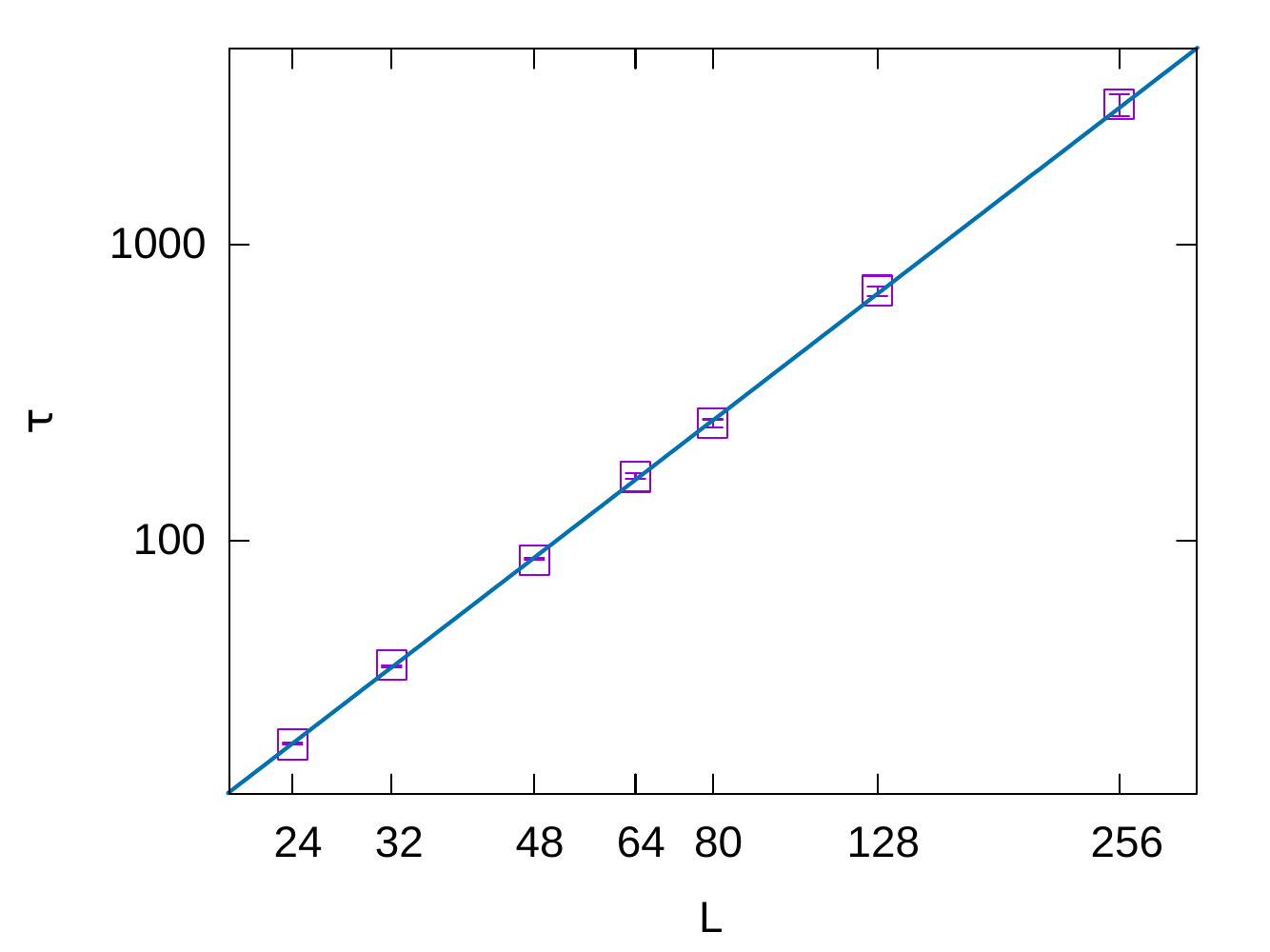}
  \caption{Log-log plots of integrated autocorrelation time of $f_\chi$ versus
  lattice size for single cluster (left) and Metropolis algorithms (right). In
  the former case, the data points corresponding to $L=24,32$ are excluded,
  resulting in an improvement of the infinite volume extrapolation.}
	\label{fig:plot-fit-cl-tau-chi}
\end{figure}

In order to assess the validity of our results, we estimate the critical dynamic
exponent $z$ according to \cref{eq:z} and compare our findings with the existing
literature. The extrapolation of $z$  for the observable $f_\chi$ is shown in
\cref{fig:plot-fit-cl-tau-chi} and the obtained values  are reported  in
\cref{tab:z}. In the extrapolations we exclude the data points corresponding to
$L=24,32$ for the cluster algorithm, based on the improvement of the infinite
volume extrapolation, that is the one of interest.
\begin{table}[t]
	\centering
	\begin{tabular}{ccc}
		Algorithm	&$z$    		& $\chi^2$   \\
		\midrule 
		Single cluster & 0.387(7) 	& 1.64\\
		Metropolis	& 2.09(1) & 0.97 \\
	\end{tabular}
	\caption{Values of $z$ related to the observable $f_\chi$, along with their associated $\chi^2$, for single cluster and Metropolis algorithms.}
	\protect\label{tab:z}
\end{table}
As for the Metropolis algorithm, we recover the value $z\approx 2$ as expected
for a local algorithm.  Regarding the single cluster algorithm, in order to
compare our results with Ref.~\cite{alg:cl2}, the following rescaling of $\tau$
is necessary:
\begin{equation*}
\tau_{1C} = \tau\dfrac{m\langle C_s \rangle}{V},
\end{equation*}
where $\langle C_s \rangle$ is the average cluster size and $m=1$ for the
two-dimensional Ising model (for further details see Ref.~\cite{alg:cluster}). By doing so, the estimated value for $z$ becomes equal to $0.137(5)$, in
agreement with Ref.~\cite{alg:cl2}.

\section{Conclusions}\label{SecIV}

The \UNEW project provides a simple yet effective tool, in an open-source environment, to address the issue of the accurate assessment of statistical
errors affecting MC-estimates of primary and derived observables.
The program implements the $\Gamma$-method, which produces better estimates of the errors with respect to the usual binning techniques by taking into account the autocorrelation present in the generated data.
The choice of \Python
as programming language aids the reuse of this software in other projects both
as a standalone executable and as a library.

We showed an application of the analysis program to the Ising model in two
dimensions.  The simulations were performed employing the
\emph{single cluster Wolff} and \emph{Metropolis} algorithms.  Our results are identical with those
obtained via \UWERR and compatible with those of the existing literature,
implying that the analysis method is well implemented in \UNEW.

\section*{Acknowledgments}

This research did not receive any specific grant from funding agencies in the
public, commercial, or not-for-profit sectors. The authors wish to gratefully
thank Marco Guagnelli for his invaluable support, precious suggestions and
feedback on the manuscript.

\appendix

\section{$\Gamma$-method for derived quantities} \label{app:1}

Here we briefly illustrate how to extend the discussion of the $\Gamma$-method
made in \cref{SecI} to the more general case of a derived quantity
$F=f(A_{\alpha})$ following Ref.~\cite{ref1:LessErr}.  Furthermore, we consider
the case of an arbitrary number $R>1$ of replica.  Accordingly, we slightly
modify the notation with respect to \cref{SecI}: we define the per-replicum
means as
\begin{equation}
\bar{a}_{\alpha}^{r}=\frac{1}{N_{r}}
\sum_{i=1}^{N_{r}} a_{\alpha}^{i,r} \label{eq:abr},
\end{equation}
whereas $\bar{\bar{a}}_\alpha$ denote the estimators for the mean values of the primary observables:
\begin{equation} \label{eq:abb}
\bar{\bar{a}}_\alpha=\frac{1}{N}\sum_{r=1}^R N_r \bar a_{\alpha}^r \; .
\end{equation}
As an estimator of the derived quantity $f$ we take
\begin{equation}
\label{fbb}
\bar{\bar{F}} = f\left(\bar{\bar{a}}_{\alpha}\right).
\end{equation}
We assume that the Taylor expansion of $f$ in the fluctuations is valid, allowing us
to write
\begin{align}
\label{eq:taylor}
\bar{\bar{F}} \simeq F
& + \sum_{\alpha} f_{\alpha}\left(\bar{\bar{a}}_{\alpha} -
A_{\alpha} \right) \nonumber \\
& + \frac{1}{2} \sum_{\alpha\beta}
f_{\alpha\beta} \left(\bar{\bar{a}}_{\alpha} -
A_{\alpha} \right)
\left(\bar{\bar{a}}_{\beta} - A_{\beta}\right),
\end{align}
where we defined
\begin{equation}
f_{\alpha} = \frac{\partial f}{\partial A_{\alpha}}, \qquad
f_{\alpha\beta} =
\frac{\partial^2 f}
{\partial A_\alpha \partial A_\beta}, \label{grad}
\end{equation}
evaluated at the exact values. The error $\sigma_F$ is related to the correlation function through
\begin{equation}
\label{cf}
\sigma_{F}^2
\simeq \frac{C_{F}}{N},
\end{equation}
where the correlation sum
\begin{equation*}
C_{F}=\sum_{\alpha\beta}f_{\alpha}f_{\beta}\sum_{t=-\infty}^{\infty}\Gamma_{\alpha\beta}(t)
\end{equation*}
generalizes \cref{eq:auto-sum}. Also in this case, similarly to \cref{eq:sigma-a}, we can rewrite
\begin{equation}
\sigma_{F}^2=\frac{2\tau_{\mathrm{int},F}}{N}v_{F}, \label{sigmaf}
\end{equation}
thus separating the contributions coming from the effective variance of $F$
\begin{equation}
\label{vf}
v_F=\sum_{\alpha\beta}f_{\alpha}f_{\beta}\Gamma_{\alpha\beta}(0),
\end{equation}
and from the integrated autocorrelation time of $F$, given by
\begin{equation}
\label{tauint}
\tau_{\mathrm{int},F}=\frac{C_F}{2v_{F}} .
\end{equation}

Exploiting the division of our data into replica, we consider as an estimator of the autocorrelation function
\begin{equation}\label{eq:gammafbb}
\bar{\bar{\Gamma}}_{\alpha\beta}(t)=\frac{1}{N-Rt}\sum_{r=1}^R\sum_{i=1}^{N_r-t}\left( a_{\alpha}^{i,r}-\bar{\bar{a}}_{\alpha}\right)\left( a_{\beta}^{i+t,r}-\bar{\bar{a}}_{\beta}\right),
\end{equation}
which reduces to \cref{eq:gamma-est} for $R=1$.
Nonetheless, one usually needs to compute only a few functions of the primary observables. Then, from a practical point of view instead of computing the whole correlation matrix it is more convenient to evaluate the projection
\begin{equation}
\label{gammafbb}
\bar{\bar{\Gamma}}_{F}(t)=\frac{1}{N-Rt}\sum_{r=1}^R\sum_{i=1}^{N_r-t}\,\delta_{F}^{i,r}\,\delta_{F}^{i+t,r},
\end{equation}
where
\begin{equation}
\label{delpro}
\delta_{F}^{i,r} = \sum_{\alpha}
\bar{\bar{f}}_{\alpha}
\left(a_{\alpha}^{i,r} - \bar{\bar a}_{\alpha}\right).
\end{equation}
Here the gradients $\bar{\bar{f}}_{\alpha}$ are defined as in \eqref{grad}, but evaluated at the mean values $\bar{\bar{a}}_{1}$, $\bar{\bar{a}}_{2},\ldots$ of the primary variables. In practice, their numerical estimates are given by the difference quotients
\begin{align*}
\bar{\bar{f}}_{\alpha} \approx
\frac{1}{2h_{\alpha}} \big[
& f\left( \bar{\bar{a}}_1, \bar{\bar{a}}_2, \ldots,
\bar{\bar{a}}_{\alpha} + h_{\alpha}, \ldots\right) \nonumber \\
& -f\left( \bar{\bar{a}}_1, \bar{\bar{a}}_2, \ldots,
\bar{\bar{a}}_{\alpha} - h_{\alpha}, \ldots \right)\big],
\end{align*}
with
\begin{equation*}
h_{\alpha}=\sqrt{\frac{\bar{\bar{\Gamma}}_{\alpha\alpha}(0)}{N}}.
\end{equation*}
For the effective variance $v_{F}$ and the error $C_{F}$ (see Eqs.~\eqref{cf} and \eqref{vf}) we take the following estimators:

\begin{equation}
\label{eq:variance}
\bar{\bar{v}}_{F}=\bar{\bar{\Gamma}}_{F}(0),
\end{equation}
\begin{equation}
\label{eq:bbc}
\bar{\bar{C}}_{F}(W)=\left[\bar{\bar{\Gamma}}_{F}(0)+2\sum_{t=1}^{W}\bar{\bar{\Gamma}}_{F}(t)\right].
\end{equation}
Notice that, as already discussed for the case of the primary observables, we need to truncate the sum in \cref{eq:bbc} to a summation window $W$. We will come back to the automatic procedure to determine the optimal value of $W$ in the following.

The estimators of the error $\sigma_F$ in Eq.~\eqref{sigmaf} and of the integrated autocorrelation time $\tau_{\mathrm{int},F}$ in Eq.~\eqref{tauint} are simply given by
\begin{equation}
\label{sigmabb}
\bar{\bar{\sigma}}_F=\sqrt{\frac{\bar{\bar{C}}_F(W)}{N}},
\end{equation}
\begin{equation}
\label{tauintfbb}
\bar{\bar{\tau}}_{\mathrm{int},F}(W)=\frac{\bar{\bar{C}}_F(W)}{2\bar{\bar{v}}_F}.
\end{equation}
It can be shown that the statistical error affecting $\bar{\bar{C}}_F(W)$ is
\begin{equation*}
\left\langle\left(\bar{\bar{C}}_F(W)-C_F\right)^2\right\rangle\approx \frac{2(2W+1)}{N}C_F^2 \, , \quad \text{for} \quad \tau\ll W \ll N,
\end{equation*}
leading to a statistical error of the error
\begin{equation}
\frac{\delta_{\text{stat}}
\left(\bar{\bar{\sigma}}_F\right)}{\bar{\bar{\sigma}}_F} =
\sqrt{\frac{W+1/2}{N}} \, . \label{ddvalue}
\end{equation}
For the error on the autocorrelation time, instead, if $\tau_{\text{int},F}\gg 1\gg e^{-W/\tau_{\text{int},F}}$, one gets
\begin{equation}
\left\langle\left(\bar{\bar{\tau}}_{\text{int},F}(W) -
\tau_{\text{int},F}\right)^2\right\rangle \approx
\frac{4}{N} \left(W + \frac{1}{2} - \tau_{\text{int},F}\right)
\tau^2_{\text{int},F}. \label{dtauint}
\end{equation}
The estimator for the derived quantity $F$ is affected by a bias due to the
truncation of the power expansion \eqref{eq:taylor}; in order to cancel it up to
the leading order $1/N$, one can exploit the division of data into replica and replace
\begin{equation}
\label{value}
\bar{\bar{F}},\bar{F}\longrightarrow \frac{R\bar{\bar{F}}-\bar{F}}{R-1},
\end{equation}
where
\begin{equation}
\label{fb}
\bar{F} = \frac{1}{N}
\sum_{r=1}^{R}\,N_r f\left(\bar{a}^r_{\alpha}\right).
\end{equation}
On the other hand, also the correlation function $\bar{\bar{\Gamma}}_F(t)$ is affected by a bias caused by the subtraction of the ensemble means $\bar{\bar{a}}_\alpha$ instead of the exact values in \cref{eq:gammafbb}. We correct this bias at leading order by first evaluating $\bar{\bar{C}}_{F}(W)$ according to \eqref{eq:bbc} and then substituting
\begin{equation*}
\bar{\bar{\Gamma}}_F(t)\longrightarrow \bar{\bar{\Gamma}}_F(t)+\frac{\bar{\bar{C}}_{F}(W)}{N}.
\end{equation*}
The new resulting $\bar{\bar{\Gamma}}_F(t)$ is eventually taken to give a second, more refined estimate of $\bar{\bar{C}}_F(W)$.

Let us now come back to the windowing procedure. Solving \cref{eq:tauw} for $\bar\tau(W)$ yields
\begin{equation*}
\frac{S}{\bar{\bar{\tau}}(W)}=\ln\left(\frac{2\bar{\bar{\tau}}_{\mathrm{int},F}(W)+1}{2\bar{\bar{\tau}}_{\mathrm{int},F}(W)-1}\right)= \frac{1}{\bar{\bar{\tau}}_{\mathrm{int},F}(W)}+\frac{1}{12\bar{\bar{\tau}}_{\mathrm{int},F}^3(W)}+\ldots,
\end{equation*}
which is valid for $\bar{\bar{\tau}}_{\mathrm{int}}(W)> 1/2$. If $\bar{\bar{\tau}}_{\mathrm{int}}(W)\leq 1/2$, instead, we set $\bar{\bar{\tau}}(W)$ to a tiny positive value. We hence evaluate, for each integer $W$, the derivative $g(W)$ (up to a factor) of Eq.~\eqref{eq:rel-err}:
\begin{equation}
g(W) = \exp\left[ -W / \bar{\bar{\tau}}(W) \right] -
\frac{\bar{\bar{\tau}}(W)}{\sqrt{WN}} \label{gw}
\end{equation}
and take as the optimal value $W_{\text{opt}}$ the first value of $W$ for which $g(W)$ becomes negative. If the windowing condition fails (\emph{i.e.} $g(W)$ does not change sign) up to $\nu= \min_r N_r / 2$, the summation window is taken as $\nu$. The correlation sum \eqref{eq:bbc} is computed only up to $t=t_{\text{max}}= 2\min(W_{\text{opt}},\nu)$.

Finally, we define some estimators which are useful for producing plots related to the derived quantity and to the autocorrelation time. The replica distribution
\begin{equation}
\label{eq:rep_dist}
p_r=\frac{f\left(\bar{a}_{\alpha}^r\right)-\bar{F}}{\sigma_F\sqrt{N/N_r-1}}
\end{equation}
is plotted, with a Q-value given by
\begin{equation}
\label{qval}
	Q = 1 - P\left(\frac{R-1}{2},\frac{x}{2}\right),
\end{equation}
where $P$ is the regularized lower incomplete Gamma function
\begin{equation*}
	P(s,y) = \frac{1}{\Gamma(s)} \int_0^y t^{s-1} e^{-t} \; \mathrm{d}t
\end{equation*}
and
\begin{equation*}
x=\sum_{r}N_{r}\,\frac{\left[f(\bar{a}_{\alpha}^r)-\bar{F}\right]^2}{\bar{\bar{C}}_{F}(W_{\text{opt}})} .
\end{equation*}
The normalized autocorrelation function
\begin{equation}
\label{rho}
\bar{\bar{\rho}}_F(t)=\frac{\bar{\bar{\Gamma}}_F(t)}{\bar{\bar{\Gamma}}_F(0)}
\end{equation}
is also plotted along with its error (given by Eqs.~(E.10) and~(E.11) of Ref.~\cite{Luscher:2005rx}).

\section{Program description} \label{app:program}

\subsection{Installation and usage} \label{sec:install}

The \UNEW package can be installed from source and the package tarball can be
downloaded from the \URL~\footnote{
	\url{http://bitbucket.org/betocollaboration/unew/get/HEAD.tar.gz}}.  \UNEW
is structured as a standard \Python package that can be installed executing the
following commands from a terminal prompt:
\begin{lstlisting}[language=bash]
$ cd /path/to/UNEW/sources
$ pip install .
\end{lstlisting}
 In case \TT{pip} is not available, one can fall back to the script \TT{setup.py} itself:
\begin{lstlisting}[language=bash]
$ cd /path/to/UNEW/sources
$ python setup.py install
\end{lstlisting}
Alternatively, one can install the package locally for a single user with the
command
\footnote{The user installation path depends on the platform and we refer to the official
	documentation for the details about the alternate ways
	of installation
	\url{http://docs.python.org/3/install/index.html\#alternate-installatio
		n}.}
\begin{lstlisting}[language=bash]
$ pip install --user .
\end{lstlisting}
The setup script manages also the installation of the required dependencies,
which are currently the \Python modules \TT{numpy}, \TT{scipy}, \TT{matplotlib}, \TT{docopt},
\TT{voluptuous}, \TT{PyYAML}, \TT{tqdm} and \TT{colorama}.

Once \UNEW is successfully installed on the target system, it can be run by
simply calling the command \TT{unew} from the command line. There are two modes of operation of the program,
depending on the ways one provides the input data files to be analysed.  In
the first case, the user provides the input files directly on the command line; in the other one, the user specifies a directory from where the input files should
be loaded. Other options can be provided to further customize the analysis. In
particular, one can perform the analysis on some derived quantity of the primary
observables, defined as \Python functions residing in a
module. For instance, suppose that a function \TT{derived} is contained in a
file named \TT{module.py}; the user needs to let the \Python interpreter
know the location of the module adding it to the environment variable
\TT{PYTHONPATH}: in \TT{bash} one can type
\begin{lstlisting}[language=bash]
$ export PYTHONPATH=/path/to/module.py:$PYTHONPATH
\end{lstlisting}
Supposing that the input data reside in the directory \TT{data}, the analysis of \TT{derived} can be issued with the following command:
\begin{lstlisting}[language=bash]
$ unew -d data -m module -q derived
\end{lstlisting}
\UNEW supports also the specification of the input parameters through a configuration file written in \YAML with the following general format:
\begin{lstlisting}[language=bash]
!UnewConfig
directory: null or /path/to/data
replica:   null or [] or [file1, file2, ...]
patterns:  null or [] or [p1, p2, ...]
indices:   null or [] or [1,3,4,7, ...]
ranges:    null or [] or [[1,10], [20,30], ...]
R:         integer (default 1)
stau:      float (default 1.5)
primaries: null or [1,2,3, ...]
params:    null or {par1: val1, par2: val2, ...}
module:    null or module_name
functions: null or [] or [func1, func2, ...]
\end{lstlisting}
Such a configuration file can be generated by adding the option \texttt{-C}.
If the file is named \TT{conf.yaml}, the program can then be run with the following command:
\begin{lstlisting}[language=bash]
$ unew -f conf.yaml
\end{lstlisting}
For further details about the usage we refer the reader to the documentation~\footnote{\url{http://bitbucket.org/betocollaboration/unew}} and for a practical example we refer to \ref{sec:example}.

\subsection{Structure of the program}

In this section, we describe in a greater level of detail the modules
introduced in \cref{sec:unew}.

The module
\ioutils provides the function \TT{load\_files} that loads the input data from a
list of file paths.  The returned object is a list of \TT{numpy} \TT{array}{s}
of replica.  If one needs to subdivide the files into more replica or in the
case there is only one file with many replica, one can pass to the function
\TT{load\_files} the parameter \TT{R} in order to split each file into $R$ replica.  The
total number of replica is then given by $R N\tp{files}$, where $N\tp{files}$ is
the number of data files given in input.
Before performing the analysis, input data are adapted through the
function \TT{prepare\_data} into a \TT{numpy} \TT{array} with dimensions $(R,
N_{r,\text{max}}, N_\alpha)$, where $N_{r,\text{max}} = \max_r N_r$.

The module \analysis is devoted to the implementation of the $\Gamma$-method.
The classes \TT{PrimaryAnalysis} and \TT{DerivedAnalysis} compute the output of
the program depending on whether the user chooses to analyse, respectively,
primary or derived observables.  In particular, the autocorrelation function is
computed using the following estimators, stored as instance variables in an
object of type \TT{AnalysisData}:
\begin{itemize}
	\item \TT{value}, the estimator of mean values, defined in \cref{eq:abb,fbb};
	\item \TT{rep\_value}, an array with the values of the derived quantity taken
	at the per-replicum means \eqref{eq:abr};
	\item \TT{rep\_mean}, the average over replica, see Eq.~\eqref{fb};
	\item \TT{deviation}, the $R\times N_{r,\text{max}}$-matrix whose entries are
	defined in Eq.~\eqref{delpro}.
\end{itemize}

The program then computes the errors and the autocorrelation times and stores the resulting data in the same \TT{AnalysisData} object; the
essential output of the program consists in the following variables:
\begin{itemize}
	\item \TT{w\_opt}, the optimal value $W_{\text{opt}}$ for the summation window $W$, see Eq.~\eqref{gw};
	\item \TT{t\_max}, the maximum value of $t$ up to which the autocorrelation function $\bar{\bar{\Gamma}}_F(t)$ \eqref{gammafbb} is computed; see the discussion following \cref{gw};
	\item \TT{value}, the unbiased expectation value of $F$ (see Eqs.~\eqref{fbb} and \eqref{value});
	\item \TT{dvalue}, the error of \TT{value}, given by Eq.~\eqref{sigmabb} with $W=W_{\text{opt}}$;
	\item \TT{ddvalue}, the statistical error of \TT{dvalue}, given by
	Eq.~\eqref{ddvalue} with $W=W_{\text{opt}}$;
	\item \TT{variance}, the variance of $F$, \eqref{eq:variance};
	\item \TT{naive\_err}, the error of $F$ computed disregarding autocorrelations, given by $\bar{\bar\sigma}_{F,\mathrm{naive}} = \sqrt{\frac{\bar{\bar v}_F}{N}}$, where $\bar{\bar v}_F$ is the variance~\eqref{eq:variance};
	\item \TT{tau\_int}, the integrated autocorrelation time, given by Eq.~\eqref{tauintfbb} with $W=W_{\text{opt}}$;
	\item \TT{dtau\_int}, the error of \TT{tau\_int}, given by Eq.~\eqref{dtauint} with $W=W_{\text{opt}}$;
	\item \TT{tau\_int\_fbb}, the partial autocorrelation times \eqref{tauintfbb};
	\item \TT{dtau\_int\_fbb}, the error of \TT{tau\_int\_fbb}, given by Eq.~\eqref{dtauint};
	\item \TT{rho}, the normalized autocorrelation function \eqref{rho};
	\item \TT{drho}, the error of \TT{rho}, see Eqs.~(E.10) and~(E.11) of~\cite{Luscher:2005rx};
	\item \TT{qval}, representing the $Q$-value for the histogram of replica distribution, given by Eq.~\eqref{qval}.
\end{itemize}

At the end of the analysis process the program draws the plots of the integrated
autocorrelation time, the autocorrelation function, the distribution of data,
and the distribution of replica. The underlying graphics engine is provided by
\TT{matplotlib}.  The plots are handled by the classes \TT{PrimaryPlot} and
\TT{DerivedPlot} which delegate the actual drawing to the class \TT{PlotHelper}:
the method \TT{autoCorrTime} plots the integrated autocorrelation time from $0$
to \TT{t\_max}, while the normalized autocorrelation function is drawn by
calling \TT{normAutoCorr}.  Additionally, the class \TT{PlotHelper} provides the
method \TT{histogram} to plot the replica distribution (see
Eq.~\eqref{eq:rep_dist}).

Furthermore, the program validates the input parameters, whether given on the
command line or through a configuration file,  and reports an error to the user
when it can not proceed with the analysis.  The validation process relies upon
the \Python module \TT{voluptuous}\footnote{\Python package for data validation,
see \url{http://pypi.python.org/pypi/voluptuous/0.9.3}.}, which presents a
simple interface and supports complex data structures.  All the
validation-related code resides in the module \configuration. This module
defines a configuration class, \TT{UnewConfig}, whose members provide all the
essential information required to perform the analysis (\emph{e.g.} data-file
names, functions, specification of the observables, number of replica). The
validation process is carried out by checking whether instances of this class
are properly constructed.  Since \UNEW can be run providing the input parameters
either as command line options or through a \YAML configuration file, the
library provides the classes \TT{CmdLineValidator} and \TT{FileValidator}
defining the appropriate validation schema in the two cases.

\subsection{Example of application} \label{sec:example}

We present an example of execution of the program for the analysis of the
derived observable $f_\chi$ as given in \cref{eq:fchi} computed with the single cluster algorithm.  In general, a derived
function needs to have a mandatory argument (the first one) representing the
mean values~\eqref{eq:abb}.  The user is free to customize the definition of the
derived function by adding an arbitrary number of arguments, such as lattice size,
critical temperature, etc. In this example the module named \texttt{ising2d}
contains the definition of the function \texttt{f\_chi}:
\begin{lstlisting}
def f_chi(abb,L):
"""
Computation of f_chi = L^(7/4)/chi,
where L is the linear lattice size
and chi is the susceptibility.
abb[3] is the mean squared magnetization and
abb[4] is the mean magnetization.
"""
V = L**2
exp = 7.0/4.0
return pow(L,exp)/((abb[3]/V-(abb[4]/V)**2))
\end{lstlisting}

\noindent The mandatory argument is here labeled as \texttt{abb}. Running the
program from the command line as
\begin{lstlisting}[language=bash]
$ unew -d /path/to/data/directory \
-m ising2d -q f_chi -P L=32 -S 1 -R 100
\end{lstlisting}
one obtains the following output:
\begingroup
\footnotesize
\begin{verbatim}
Results for derived quantity 'f_chi':
         value: 9.164948766077298e-01
         error: 2.945798278639178e-04
error of error: 4.885053799160480e-07
   naive error: 1.183870408242084e-04
      variance: 1.401549143511278e-01
       tau_int: 3.095761227344753e+00
 tau_int error: 9.672307277108475e-03
         W_opt: 27
         t_max: 54
         Q_val: 8.919403963566213e-01
\end{verbatim}
\endgroup
\noindent Equivalently, one can execute the program with the command line
\begin{lstlisting}[language=bash]
$ unew -f config.yaml
\end{lstlisting}
where the file \texttt{config.yaml} is the configuration file:
\begin{lstlisting}[language=bash]
!UnewConfig
R: 100
directory: /path/to/data/directory
functions: [f_chi]
indices: null
module: ising2d
params: {L: 32.0}
patterns: []
primaries: null
ranges: []
replica: []
stau: 1.0
\end{lstlisting}

\begin{figure}[t]
	\centering
    \includegraphics[scale=0.5]{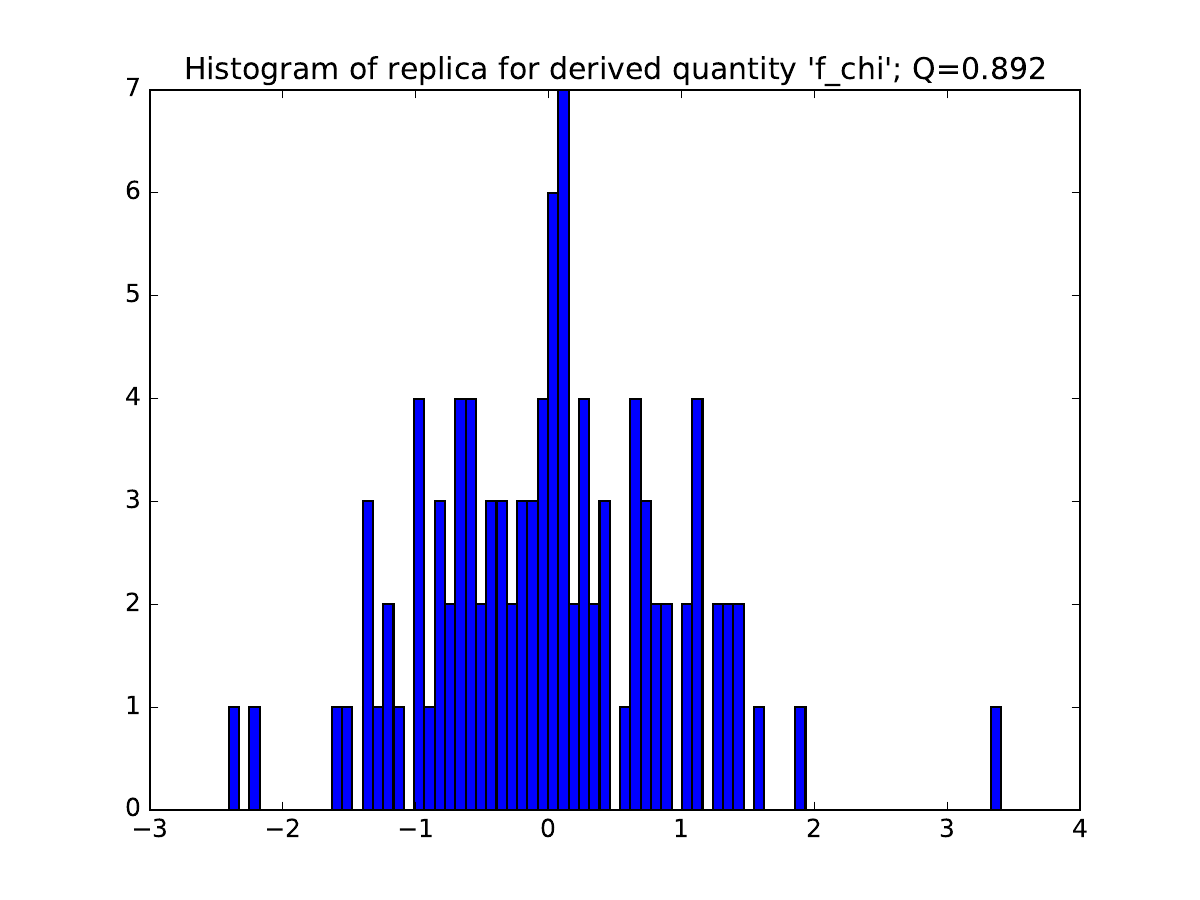}
	\caption{Distribution of replica ($R=100$) for the derived function $f_\chi$.}
	\label{fig:fchi_hist}
\end{figure}

Additionally, the program produces the plots in \cref{fig:tau_rho} and the
histogram of replica as shown in \cref{fig:fchi_hist}.


\section*{References}
\bibliographystyle{plainnt}
\bibliography{bib}

\begin{thebibliography}{1}

\bibitem{Luscher:2005rx}
M.~Luscher.
\newblock {\em Comput. Phys. Commun.}, 165:199--220, 2005.

\bibitem{Madras1988}
N.~Madras and A.D. Sokal.
\newblock {\em J. Stat. Phys.}, 50(1):109--186, 1988.

\bibitem{alg:metro}
N.~Metropolis, A.W. Rosenbluth, M.N. Rosenbluth, A.H. Teller, and E.~Teller.
\newblock {\em J. Chem. Phys.}, 21(6):1087--1092, 1953.

\bibitem{Priestley1981}
M.~B. Priestley.
\newblock {\em Spectral analysis and time series / M.B. Priestley}.
\newblock Academic Press London; New York, 1981.

\bibitem{alg:cluster}
U.~Wolff.
\newblock {\em Phys. Rev. Lett.}, 62:361--364, Jan 1989.

\bibitem{ref1:LessErr}
U.~Wolff.
\newblock {\em Comput. Phys. Commun.}, 156:143--153, 2004.
\newblock Erratum-ibid.~\cite{ref:erratum}.

\bibitem{ref:erratum}
U.~Wolff.
\newblock {\em Comput. Phys. Commun.}, 176(5):383, 2007.

\bibitem{alg:cl2}
Ulli Wolff.
\newblock {\em Physics Letters B}, 228(3):379 -- 382, 1989.

\end{thebibliography}

\end{document}